\documentstyle[12pt]{article}
\begin{document}
\begin{center}
{\large \bf LORENTZ SYMMETRY VIOLATION} \\[0pt] \vspace{0.1cm} 
{\large \bf AND VERY HIGH-ENERGY CROSS SECTIONS}\\[0pt]
\vspace{0.5cm} {\bf Luis GONZALEZ-MESTRES}\footnote{%
E-mail: lgonzalz@vxcern.cern.ch}
\\[0pt]
\vspace{0.3cm}
{\it Laboratoire de Physique Corpusculaire, Coll\`ege de France \\
11 pl.
Marcellin-Berthelot, 75231 Paris Cedex 05, France
\\[0pt] and
\\[0pt]
Laboratoire d'Annecy-le-Vieux de Physique des Particules \\ B.P. 110 , 74941 
Annecy-le-Vieux Cedex, 
France} \vspace{0.8cm}
\end{center}

\begin{abstract}

We discuss the implications of a recently proposed pattern of Lorentz 
symmetry violation on very high-energy cross sections. As a consequence of 
the breaking of local Lorentz invariance
by the introduction of a fundamental length, $a$ , 
the kinematics is modified and 
the properties of final states are 
fundamentally different in collider-like (two incoming particles with 
equal, opposite
momenta with respect to the vacuum rest frame) and fixed-target 
(one of the incoming particles at rest with respect to 
the vacuum rest
frame) situations. In the first case, the properties 
of the allowed final states 
are similar to relativistic kinematics, as long as the 
relevant wave vectors are much smaller than the critical wave vector scale
$a^{-1}$ .
But, if one of the incoming particles is close to rest in the
vacuum rest frame, energy conservation reduces the final-state
phase space at very high energy and can lead to a sharp fall of cross sections
starting at incoming-particle wave vectors well below the 
inverse of the fundamental
length. Then, the Froissart bound may cease to be relevant, as total cross 
sections seem to become much smaller than it would be allowed
by local, Lorentz-invariant, field
theory. Important experimental implications of the new scenario are found 
for cosmic-ray astrophysics
and for very high-energy cosmic rays reaching the earth.  
 
\end{abstract}

\section{Introduction}

In two previous papers (Gonzalez-Mestres, 1997a and 1997b), we suggested
that, as a consequence of nonlocal dynamics at Planck scale or at some other
fundamental length scale, Lorentz symmetry violation can result in a
modification of the equation relating energy and momentum which
would write in the vacuum rest frame:
\equation
E~~=~~(2\pi )^{-1}~h~c~a^{-1}~e~(k~a)
\endequation
where $E$ is the energy of the particle,
$h$ the Planck constant, $c$ the speed of light,
$a$ a fundamental length scale (that we can
naturally identify with the Planck length, but other choices of the
fundamental distance scale are possible), $k$
the wave vector modulus and
$[e~(k~a)]^2$ is a convex
function of $(k~a)^2$ obtained from nonlocal vacuum dynamics.

Rather generally, we find
that, at wave vector scales below the inverse of the fundamental length scale,
Lorentz symmetry violation in relativistic kinematics can be parameterized
writing:
\equation
e~(k~a)~~\simeq ~~[(k~a)^2~-~\alpha ~(k~a)^4~+~(2\pi ~a)^2~h^{-2}~m^2~c^2]^{1/2}
\endequation
where $\alpha $ is a positive
constant between $10^{-1}$ and $10^{-2}$ . At high energy, we can write:
\equation
e~(k~a)~~\simeq ~~k~a~[1~-~\alpha ~(k~a)^2/2]~
+~2~\pi ^2~h^{-2}~k^{-1}~a~m^2~c^2
\endequation
and, in any case, we expect observable kinematical effects when the term
$\alpha (ka)^3/2$ becomes as large as the term
$2~\pi ^2~h^{-2}~k^{-1}~a~m^2~c^2$ .
Assuming that, apart form the value of the mass, expression (2) is universal
for all existing particles whose critical speed in vacuum is equal to the speed 
of light in the Lorentz-invariant limit, we found three important efects:

a) The Greisen-Zatsepin-Kuzmin (GZK) cutoff on very high-energy cosmic
protons and nuclei (Greisen, 1966; 
Zatsepin and Kuzmin, 1966) does no longer apply.

b) Unstable particles with at least two massive particles in the final state
of all their decay channels become stable at very high energy.

c) In any case, unstable particles live longer than naively expected with exact
Lorentz invariance and, at high enough energy,
the effect becomes much stronger than previously estimated for nonlocal models
(Anchordoqui, Dova, G\'omez Dumm
and Lacentre, 1997)
ignoring the small violation of relativistic kinematics.

Furthermore, velocity reaches its maximum at 
$k~\approx ~(4\pi ^2~\alpha ^{-1}/3)^{1/4}~(m~c~h^{-1}~a^{-1})^{1/2}$ .
Above this value, increase of momentum amounts to deceleration.
In our ansatz, 
observable effects of local Lorentz invariance breaking
arise, at leading level, well below the critical
wavelength scale $a^{-1}$ due to the fact that, contrary to previous models 
(f.i. R\'edei, 1967), we directly apply non-locality to particle 
propagators and not only 
to the interaction hamiltonian. In contrast with 
previous
patterns (f.i. Blokhintsev, 1966), $s-t-u$ kinematics ceases to make sense
and the motion of the global system with respect to the vacuum rest frame plays
a crucial role. The physics of elastic two-body scattering will depend  
on five kinematical variables.
Noncausal dispersion relations (Blokhintsev and Kolerov, 
1964) should be reconsidered, taking into account the departure from 
relativistic kinematics. 

In this note, we would like to discuss another important consequence of the
new kinematics, i.e. the appearence of strong limitations in the allowed
phase space for final states of two-body collisions,
especially when the target is moving
slowly with respect to the vacuum rest frame. As in previous papers
(Gonzalez-Mestres, 1997a and 1997b), we assume that 
$c$ and $\alpha $ are universal
constants for all particles under consideration. If this were not the case,
our analysis would require modifications but other new physical phenomena 
would equally emerge. Such an alternative will be discussed in a forthcoming
paper.


\section{The new kinematics}

No special constraint seems to arise from (2) if, in the vacuum rest frame, two
particles with equal, 
opposite momenta of modulus $p$ with $\alpha ~(k~a)^2~
\ll ~1$ collide to produce a multiparticle final state. 
When the term
$\alpha ~(k~a)^2~p~c/2$ becomes $\approx m^2~c^2~p^{-1}/2$ or larger,
the new kinematics favours large momenta 
and allows for new final-state phase space, as compared
to relativistic kinematics. 
But, as a consequence
of  
Lorentz symmetry violation
(the required transformation would have relative speed $v~\simeq ~c$), 
the situation becomes fundamentally different 
at very high energy if
one of the incoming 
particles is close to rest with respect to the "absolute" frame
where formulae (1)~-~(3) apply.

Assume a very high-energy particle (particle 1) 
with momentum ${\vec {\bf p}}$ , impinging on a particle
at rest (particle 2) in the vacuum rest frame. We take  
both particles to have mass $m$ , and $p~\gg ~mc$ .
In relativistic kinematics, we would have elastic final states where particle
1 has, with respect to the direction of ${\vec {\bf p}}$ ,
longitudinal momentum $p_{1,L}~\gg ~mc$ and particle 2 has 
longitudinal momentum
$p_{2,L}~\gg ~mc$  with $p_{1,L}~+~p_{2,L}~=~p$ . A total transverse energy 
$E_T~\simeq ~mc^2$ would still be left for the outgoing particles. 
However, the situation is drastically modified if the kinematics is given
by expressions (1) - (3) and if $\alpha ~(k~a)^2~p$ ($k$ being the 
modulus of the wave vector
of the incoming particle)
becomes of the same order 
as $m~c$ or larger. 
As
the energy increases, stronger and stronger limitations of the available
final-state phase space appear:
with the 
approximation 
(3),   
the final-state configuration $p_{1,L}~=~p~-~p_{2,L}~=~(1~-~\lambda )~p$
becomes kinematically forbidden for $\alpha ~(k~a)^2~p~>~2~m~c~\lambda ^{-1} 
(1~-~\lambda )^{-1}/3$ . Thus, for momenta above $\approx 
(m~c~a^{-2}~h^2)^{1/3}$ , 
"hard" interactions become severely
limited by
kinematical constraints. 

Similarly, with the same initial state,
a multiperipheral final state configuration with $N$ particles ($N~>~2$)
of mass $m$ and longitudinal momenta $g^{i-1}~p'_L$ 
($i~=~1,...,N$ , $g~>
 ~1$), 
where $p'_L~=~p~(g~-~1)~(g^{N}~-~1)^{-1}$ , $g^N~\gg ~1$ 
and $p'_L~\gg m~c$ , would have in 
standard relativity an allowed total transverse energy
$E_T~(N~,~g)~\simeq ~
m~c^2~[1~-~m~c~(2~p'_L)^{-1}~(1~-~g^{-1})^{-1}]$ which is positive definite.
Again, using the new kinematics and the approximation (3), we find that
such a longitudinal final state configuration is forbidden for 
values of the incoming momentum such that 
$\alpha ~(k~a)^2~p~c~>~2~(3~g)^{-1}~(1~+~g~+~g^2)~E_T~(N~,~g)$ .

The above, or similar, considerations apply to strong interactions as well as
to electromagnetic processes.
For the initial state configuration
where the target is at rest in the vacuum rest frame, 
and compared to standard expectations based on relativistic
kinematics,
a {\it sharp fall of elastic, multiparticle and total cross sections} 
can be expected at very high energy.
For "soft" strong interactions, the approach were the two-body
total cross section is the less sensitive to final-state phase space
is, in principle, that based on dual resonance models and considering the
imaginary part of the elastic amplitude as being dominated by the shadow
of the production of pairs of very heavy resonances of masses $M_1$ and $M_2$
of order $\approx (p~m~c^3/2)^{1/2}$ in the direct channel
(Aurenche and Gonzalez-Mestres, 1978 and 1979). 
But, even in this scenario, we find important limitations to the
allowed values of $M_1$ and $M_2$ , and to the two-resonance phase space, 
when $\alpha ~(k~a)^2~p$ becomes $\approx m~c$ or larger.
In all cases, the departure from the standard relativistic situation 
occurs, if the target is close to rest in the vacuum rest frame, 
at incoming energies
$E$ above $\approx (m~a^{-2}~h^2~c^4)^{1/3}$ 
which corresponds to a transition energy scale $\approx 10^{22}~eV$
for $m~\approx~1~GeV/c^2$ and
$a~\approx ~10^{-33}~cm$~,  
and $\approx 10^{21}~eV$ 
if the target mass is $\approx 500~keV/c^2$ . 
Lowering the critical wave vector scale $a^{-1}$ to 
$~\approx ~10^{26}~cm^{-1}$
(just above the wave vector scale of the highest-energy cosmic
rays), the fall of cross sections would start at 
$E~\approx 10^{16}~-~10^{17}~eV$ , which seems excluded by cosmic ray
data
if the earth is moving slowly with respect to the vacuum rest frame.
In astrophysical processes,
the new kinematics may inhibit phenomena such as GZK-like cutoffs, 
photodisintegration of nuclei, decays, 
radiation emission under external forces, momentum loss 
(which at very high energy does not imply deceleration) through collisions,
production of lower-energy secondaries... {\it potentially solving the basic 
problems raised by the highest-energy cosmic rays}. 
Above $E~\approx (m~a^{-2}~h^2~c^4)^{1/3}$~, nonlocal
effects play a crucial role and invalidate considerations based 
on Lorentz invariance and local field theory used to derive the Froissart
bound (Froissart, 1961), which seems not to be violated but ceases
to be significant given the expected behaviour of total 
cross sections which, at very high-energy, 
seem to fall far below this bound. 
An updated study of noncausal dispersion relations, 
incorporating 
the new kinematics from nonlocal dynamics, can 
possibly lead new bounds. As previously stressed (Gonzalez-Mestres,
1997a) , this apparent
nonlocality may actually reflect the existence of superluminal sectors of 
matter (Gonzalez-Mestres, 1996) where causality would hold at the 
superluminal level (Gonzalez-Mestres, 1997c).

Other initial state configurations can be considered. 
We may have two incoming particles with momenta of moduli $p_1^i$
and $p_2^i$ and opposite directions in the vacuum rest frame, and $p_1^i~
\gg ~p_2^i~\gg ~mc$ . Keeping a constant value of $\lambda ~=~p_2^i~
(p_1^i)^{-1}$ , we find 
that the fall of final-state
phase space occurs for $p_1^i$ above $\approx \lambda ^{1/2}~a^{-1}~h$~. 
The incoming momenta $p_1^i$
and $p_2^i$ may also be pointing in the same direction.
Then, the final-state
phase space starts to fall at $p_1^i~\approx 
\lambda ^{-1/4}~(m~c~h~a^{-1})^{1/2}$ . A more complete discussion, 
including non-parallel incoming momenta and the case $m~=~0$ , will
be presented elsewhere.

\section{Experimental considerations}

Lorentz symmetry violation
prevents naive extrapolations from reactions between two particles 
with equal, opposite momenta in the vacuum rest frame 
(similar to colliders) to reactions where 
the target is at rest in this frame
(similar to cosmic-ray events). {\it Assuming the earth to move slowly with
respect to the vacuum rest frame} (for instance, if the "absolute" frame
is close to that defined by the requirement of cosmic microwave background
isotropy),
the described kinematics predicts the existence of 
a maximum energy deposition for high-energy
cosmic rays in the atmosphere, in the rock or in a given underground or
underwater detector. Well below Planck energy, a very high-energy
cosmic ray would not necessarily
deposit most of its energy in the atmosphere: its energy 
deposition decreases for energies above a transition 
scale, far below 
the energy scale associated to the fundamental length.
The maximum allowed
momentum transfer in a single collision occurs at an energy just below
$E~\approx (m~a^{-2}~h^2~c^4)^{1/3}$~.
For $E$ above $\approx (m~a^{-2}~h^2~c^4)^{1/3}$ , 
the allowed longitudinal momentum transfer falls, typically, 
like $p^{-2}$ 
(obtained differentiating
the term $\alpha ~k^2~a^2~p~c/2$). 
To set upper limits, 
we can take for $m$ the mass of oxigen or nitrogen in the case of 
air, oxigen in 
water, and heavier elements in the rock.  At energies around 
$\approx (m~a^{-2}~h^2~c^4)^{1/3}$ , 
the cosmic ray will in our scenario undergo several 
scatterings in the atmosphere and still lose there most of its energy,
possibly
leading to unconventional longitudinal cascade development profiles
that could be observed by very large-surface air shower detectors
like the AUGER observatory (AUGER Collaboration, 1997).
Above $E~ \approx (m~a^{-2}~h^2~c^4)^{1/3}$~, it can indeed cross the
atmosphere keeping most of its momentum and energy and deposit its energy 
in the rock or in water, or possibly reach and underground or underwater
detector. Thus, some cosmic ray events of apparent energy far below 
$10^{20}~eV$ (perhaps apparently muon or neutrino-like, or exotic-like), 
as seen by earth-surface
(e.g. air shower), underground or underwater detectors,
may actually be originated by extremely-high energy cosmic rays well above
this energy scale. 

Interesting constraints on the fundamental length $a$ can be derived
from this analysis, assuming simultaneoulsy (Gonzalez-Mestres, 
1997a and 1997b) 
that the absence of
GZK cutoff is due to the same pattern of Lorentz symmetry violation.
The
combined absence of GZK cutoff and existence of $\approx 10^{20}~eV$ energy
deposition from cosmic rays in the atmosphere 
lead to $a$ in the range $10^{-35}~cm~<~a~<~10^{-30}~cm$ 
(energy scale between $10^{16}$ and $10^{21}$ $GeV$). The lower bound comes
from the requirement that the violation of local Lorentz invariance at the
fundamental length scale be able to influence particle interactions
at the $10^{19}~-~10^{20}~eV$ energy scale strongly
enough to suppress the GZK cutoff. The
upper bound is derived from the existence of events with
$\approx 10^{20}~eV$ energy deposition in the atmosphere (Linsley, 1963;
Lawrence, Reid and Watson, 1991; Afanasiev et al., 1995; 
Bird et al., 1994; Yoshida et al., 1995).

Then, very high-energy accelerator and cosmic-ray experiments would indeed
be complementary research lines: the results
of both kinds of experiments would not be equivalent up to Lorentz 
tranformations.
If the transition energy scale for cross-sections corresponds to 
$p_1^i~c~\approx 10^{20}~eV$ ,
a  
$p~-~p$ collider at $\approx 700~TeV$ per beam
could make possible direct tests of
Lorentz symmetry violation, comparing collisions at the accelerator with 
collisions between a $\approx 10^{21}~eV$ proton 
of cosmic origin and a proton or nucleus from the
atmosphere. Simultaneously, other kinds of tests may be possible through the
lifetimes and decay products of very high-energy unstable particles  
(Gonzalez-Mestres, 1997a and 1997b) in the cosmic-ray events 
producing the highest-energy secondaries. 
We would be confronted to a new situation, 
contrary to previous expectations, if the cosmic rays at the highest possible
energies interact more and more weakly with matter because of kinematical
constraints. The existence of a maximum energy of events generated in
the atmosphere would not correspond to a maximum energy of 
incoming cosmic rays.
Unconventional events   
originated by such particles 
may have been erroneously 
interpreted as being associated to cosmic rays of much lower energy.
New analysis seem necessary, as well as 
new experimental designs 
using perhaps in coincidence very large-surface 
detectors devoted to 
interactions in the atmosphere with 
very large-volume underground or undewater detectors.
\vskip 6mm
{\bf Acknowledgement}
\noindent
\vskip 4mm
It is a pleasure to thank P. Espigat and other colleagues at LPC, Coll\`ege
de France, for useful discussions.  
\vskip 6mm
{\bf References}
\vskip 4mm
\noindent
Afanasiev, B.N. et al., Proc. of the $24^{th}$ International Cosmic Ray 
Conference, Rome, Italy, Vol. 2 , p. 756 (1995).\newline
Anchordoqui, L., Dova, M.T., G\'omez Dumm, D. and Lacentre, P.,
{\it Zeitschrift f\"{u}r Physik C} 73 , 465 (1997).\newline
AUGER Collaboration, "The Pierre Auger Observatory Design Report"
(1997).\newline
Aurenche, P. and Gonzalez-Mestres, L., {\it Phys. Rev. D} 18 , 2995 (1978).
\newline
Aurenche, P. and Gonzalez-Mestres, L., {\it Zeitschrift f\"{u}r Physik C} 1 ,
307 (1979).\newline
Bird, D.J. et al., {\it Ap. J.} 424 , 491 (1994).\newline 
Blokhintsev, D.I. and Kolerov, G.I., {\it Nuovo Cimento} 
34 , 163 (1964).\newline
Blokhintsev, D.I., {\it Sov. Phys. Usp.} 9 , 405 (1966).\newline
Froissart, M., {\it Phys. Rev.} 123, 1053 (1961). \newline
Gonzalez-Mestres, L., "Physical and Cosmological Implications of a Possible
Class of Particles Able to Travel Faster than Light", contribution to the
28$^{th}$ International Conference on High-Energy Physics, Warsaw July 1996 .
Paper hep-ph/9610474 of LANL (Los Alamos) electronic archive (1996).\newline
Gonzalez-Mestres, L., "Vacuum Structure, Lorentz Symmetry and Superluminal
Particles", paper physics/9704017 of LANL 
electronic archive (1997a).\newline
Gonzalez-Mestres, L., "Absence of Greisen-Zatsepin-Kuzmin Cutoff 
and Stability of Unstable Particles at Very High Energy,
as a Consequence of Lorentz Symmetry Violation", paper 
physics/9705031 of LANL electronic archive (1997b).\newline
Gonzalez-Mestres, L., "Space, Time and Superluminal Particles",
paper physics/9702026 of LANL electronic archive (1997c).\newline
Greisen, K., {\it Phys. Rev. Lett.} 16 , 748 (1966).\newline
Lawrence, M.A., Reid, R.J.O. and Watson, A.A., {\it J. Phys. G }, 17 , 773
(1991).\newline
Linsley, J., {\it Phys. Rev. Lett.} 10 , 146 (1963).\newline
R\'edei, L.B., {\it Phys. Rev.} 162 , 1299 (1967).\newline
Yoshida, S. et al., Proc. of the $24^{th}$ International Cosmic Ray 
Conference, Rome, Italy, Vol. 1 , p. 793 (1995).\newline
Zatsepin, G.T. and
Kuzmin, V.A., {\it Pisma Zh. Eksp. Teor. Fiz.} 4 , 114 (1966).
\end{document}